\documentstyle[amsfonts,epsfig,11pt]{article}

\def\mypagenumber{1}

\def\myend{\end{document}}

\normalsize

\newcounter{sxn}

\newcounter{axn}

\date{}

\newdimen\mybaselineskip
\newcommand{\beeq}{\begin{equation}}
\newcommand{\eneq}{\end{equation}}
\newcommand{\be}{\begin{eqnarray}}
\newcommand{\ee}{\end{eqnarray}}
\newcommand{\bpic}{\begin{picture}}
\newcommand{\epic}{\end{picture}}

\def\la{\raise.16ex\hbox{$\langle$} \, }
\def\ra{\, \raise.16ex\hbox{$\rangle$} }

\def\psibar{ \psi \kern-.65em\raise.6em\hbox{$-$} }
\def\mbar{ m \kern-.78em\raise.4em\hbox{$-$}\lower.4em\hbox{} }

\def\n@space{\nulldelimiterspace=0pt \mathsurround=0pt }
\def\huge#1{{\hbox{$\left#1\vbox to 20.5pt{}\right.\n@space$}}}

\def\myskip{\noalign{\kern 8pt}}
\def\myeqspace{\noalign{\kern 10pt}}

\def\boxit#1{$\vcenter{\hrule\hbox{\vrule\kern3pt
    \vbox{\kern3pt\hbox{#1}\kern3pt}\kern3pt\vrule}\hrule}$}
\def\bigbox#1{$\vcenter{\hrule\hbox{\vrule\kern5pt
     \vbox{\kern5pt\hbox{#1}\kern5pt}\kern5pt\vrule}\hrule}$}

\def\ignore#1{{}}

\begin{document}

\bibliographystyle{unsrt}
\footskip 1.0cm

\thispagestyle{empty}
\setcounter{page}{\mypagenumber}

%{\baselineskip=10pt \parindent=0pt \small
%\mydate
%}

\begin{flushright}{UFIFT-HEP-02-37\\}
\end{flushright}
\begin{flushright}{BRX-TH-512\\}
\end{flushright}

\vspace{2.5cm}
\begin{center}
{\LARGE \bf {Kaluza-Klein Monopole in AdS Spacetime}}\\
\vspace{1cm} {\large Vakif K. Onemli$^{a,}$\footnote{e-mail:~
onemli@phys.ufl.edu}, \hskip 0.3 cm Bayram
Tekin$^{b,c}$\footnote{e-mail:~
tekin@brandeis.edu} }\\
\vspace{.5cm}
$^a${\it Physics Department, University of Florida, Gainesville,
FL 32611, USA}\\
\vspace{.5cm}
$^b${\it {Department of Physics, Brandeis University , Waltham, MA 02454,
USA}}\\
\vspace{0.5cm} $^c${\it{Physics Department, Middle East Technical
University, 06531 Ankara, Turkey}}
\end{center}

\vspace*{1cm}
%\baselinestretch{2.0}

%\normalsize

\begin{abstract}
\baselineskip=18pt
We construct analogs of the flat space Kaluza-Klein (KK) monopoles
in locally Anti-de Sitter (AdS) spaces for $D \ge 5+1$.  We show
that, unlike the flat space KK monopole,
there is no five dimensional static KK monopole in AdS
that smoothly reduces to the flat space one
as the cosmological constant goes to zero.
Thus, one needs at least two extra
dimensions, one of which is compact, to get a static KK monopole
in cosmological backgrounds.
\end{abstract}
\vfill

Keywords: ~ Kaluza-Klein Monopole, Solitons, AdS

%\end{titlepage}

\newpage

%\setcounter{page}{1}

%\textheight=20cm
%\headsep=0.75cm
%\vsize=20cm

%%%%%%%%%%%%%%%%%%%%%%%%%%%%%%%%%%%%%%%%%%%%%%%%%%%%%%%%%%%%%%%%%
\normalsize
\baselineskip=20pt plus 1pt minus 1pt
\parindent=24pt
%\vspace*{5mm}

\section{Introduction}

Gravitational solitons are topologically stable, everywhere
smooth, particle-like solutions of pure gravity theories. Not
all dimensions accommodate these spacetimes. For example,
$D=4$ General Relativity without
a cosmological constant does not have solitons and thus,
globally flat (Lorentzian) $R^{(3,1)}$
is the unique singularity free solution. This no-go result
in four dimensions persists even if a cosmological constant ($\Lambda$)
is added to the Einstein's theory. As shown in \cite{boucher},
if the soliton is required to be asymptotically AdS ($\Lambda <0$),
then globally AdS spacetime is the unique, static, non-singular solution.
For de-Sitter ($\Lambda >0$) case, the cosmological horizon
complicates a rigorous proof, but
a similar no-soliton result is expected to hold \cite{boucher}.

Non-trivial soliton solutions
can be found if either one
or both of the following requirements are relaxed:\\
a) the spacetime is four dimensional,\\
b) the solutions  are asymptotically flat or asymptotically AdS.\\
In fact, adding an extra compact spatial dimension leads to
many solitons one of which is the well-known  Kaluza-Klein monopole
of Sorkin, Gross and Perry \cite{sorkin, gross}.
This is a $4+1$ $D$, static,
Ricci-flat space-time that looks exactly like a Dirac monopole
to a $3+1$ $D$ `effective' observer who cannot see
the compact extra dimension. Magnetic field of the
KK monopole spreads radially in the $3 D$
space. Once the size of the extra dimension is fixed, to get
the proper $3+1$ $D$ gravity plus Maxwell's Electromagnetic
theory, monopole's
{\it{inertial mass}} is determined to be  about three times the Planck
mass ( $6 \times 10^{-5}$ g ) \cite{gross}. In spite of its finite
mass, the monopole does not exert
any gravitational force on massive neutral test particles. It only interacts
with the moving charged particles, as a monopole should do.
Also, in asymptotically {\it{locally}} AdS spacetimes (as opposed
to condition (b) ),
a non-trivial solution (the AdS soliton) was
found  more recently by Horowitz and Myers \cite{horowitz}.
The AdS soliton, which exists even in four dimensions, is conjectured to
have the minimum (negative) energy among all the asymptotically locally
AdS spacetimes.
The uniqueness of the AdS soliton as the lowest
energy configuration provided support for a new `positive energy
conjecture' in gravity \cite{galloway}.

Although the Ricci flat KK monopoles are well studied
both in pure gravity theories and in string/M-theory \cite{townsend},
analogous solutions in cosmological backgrounds have not been
constructed before. In this paper, we show that in the $D=4+1$
cosmological
Einstein theory, such `AdS KK monopole' solutions that smoothly
goes to the flat space KK monopole as $\Lambda\rightarrow 0$ do not exist.
[Here, we insist on recovering the flat space solutions
as $\Lambda \rightarrow 0$. Since we are specifically looking for the
`KK monopole' in a background with negative cosmological constant, we
should obtain the flat monopole metric in this limit. ]
 However, for $D \ge 5+1$, we find static KK monopole-like solitons with
asymptotically locally AdS geometries. Given the recent interest
in the AdS spacetime both in the context of AdS/CFT duality and the brane
world scenarios, it seems proper to study monopoles in this background.

The outline of the paper is as follows: In Sec. 2, we review the flat
space KK monopole and show that there is no static monopole solution in $5D$
cosmological backgrounds.
In Sec. 3, we construct a KK monopole in $D=6$ asymptotically locally AdS
background; find another new soliton in this dimension and interpret
it as a brane-world. We conclude with Sec 4.

\section{KK monopole and the no-go result in $5D$ AdS}

First, let us review the essence of the Ricci-flat KK monopole
construction \cite{sorkin, gross}, and try to formulate a recipe for the
AdS case. Our conventions are: $(-,+,+, ... +)$ for the signature,
$\left[\nabla_M, \nabla_N\right]V_L = R_{M N L}\,^S V_S $,\,\,
$ R_{M N} \equiv R_{M L N}\,^L$ for the Riemann and Ricci
curvatures, respectively. The flat KK monopole was obtained by
trivially adding a time direction to the four dimensional Taub-NUT (TN)
space:
\be
ds^2= -dt^2  + ds^2_{\mbox{{\scriptsize TN}}}\; ,
\label{themetric}
\ee
where
\be
ds^2_{\mbox{{\scriptsize TN}}}
=V^{-1}(r)\,
\left(dx^5 + \vec{A}\cdot d\vec{r}\right)^2 +
V(r)d\vec{r}\cdot d \vec{r} \; .
\label{metric1}
\ee
TN metric is a gravitational instanton: a
Euclidean, Ricci-flat metric
that solves the $D=4$ self-dual Einstein's equations which reduce
to a first order BPS-like equation:
\be
\vec{\nabla} \times \vec{A}=\pm \vec{\nabla} V(r)\; .
\label{BPS}
\ee
Since the time direction is flat,
the $5D$ metric is also Ricci flat. The metric has a
time-like Killing vector, along with $g_{0i}=0 $. The BPS equation
implies that $V(r)$ satisfies the harmonic equation on $R^3$. Charge-one
monopole solution of (\ref{BPS}) with a $\delta$-function
singularity is:
\be
V = 1 + {4 M\over r}\;.
\ee
Thus, the vector potential $\vec{A}$ and the magnetic field $\vec{B}(r)$
are:
\be
\vec{A}(r, \theta)=
\frac{4M\left(1-\cos{\theta}\right)}{r\sin{\theta}}\hat\phi\; , \hskip 1
cm
\vec{B}(r) = {4M \hat{r}\over r^2}\; .
\label{potential}
\ee
Metric ({\ref{metric1}) becomes regular everywhere
if  $x^5$ is compactified on a circle with radius $8M$.
This solution cannot decay to the trivial vacuum (zero charge, $M=0$
case)
because of its non-trivial topological structure
with 1 unit of Euler character $(\chi_E)$. For our purposes, it is
important to notice that in the zero magnetic charge limit, as
$M\rightarrow 0$, $V(r)\rightarrow 1$ and $\vec{A}(r)\rightarrow 0$. Thus,
flat KK monopole metric (\ref{themetric}) becomes the $5 D$ Minkowski
metric:
\be
ds^2=-dt^2+dx^2_4+d\vec{r}\cdot d\vec{r} \;\; .
\ee
Therefore, the interpretation is that, when the charge $M$ is turned on
in a $5 D$ Minkowski spacetime, the monopole curves the background so that
the metric takes the form of (\ref{themetric}).

Multi-monopole solutions with equal charges (both
in sign and in magnitude ) located at various points  can also be
obtained,
since there are no forces between these BPS objects \cite{gibbons}.
In \cite{onemli}, it was shown that `KK vortices', with magnetic fields
trapped effectively in $2D$, can be constructed by summing up infinitely
many
KK monopoles on a line. Also, by adding 5 or 6 more spatial flat
directions
to the metric, one gets a $D5$ or $D6$ brane which is a BPS solution to
string or M theory, respectively \cite{townsend}.

How can one generalize the above construction to
the spaces with constant non-zero curvature?
First, there is the issue of the sign of the cosmological
constant. Since the spatial hypersurfaces of de-Sitter space (dS)
are compact (with no boundary),
one should not expect a single monopole solution in dS. Rather,
magnetic dipole-like  totally neutral
solutions might exist. [ Related is the issue of the
non-existence of a global time-like Killing vector in the relevant
solutions in dS.]
On the other hand, in AdS which has open hypersurfaces, and hence a
boundary to define flux,
one expects monopole solutions. However,
this expectation will turn out to be wrong in $D=5$. Unlike the
flat space case, for negative
cosmological backgrounds, $D=6$ is the minimal number of dimensions
where a monopole-like solution exists for pure gravity equations.

Now, let us adapt the essential ingredients of the above
Ricci flat construction to AdS. The first thing we need is a proper
background metric of a spacetime with
a negative cosmological constant $\Lambda$ (i.e. the zero charge
monopole). Needless to
say that, in this case,
the time direction cannot simply be added to a four dimensional gravitational
instanton such as the Taub-NUT or the AdS-Taub-NUT.
It has to be a warped
product metric. For $\Lambda := -2 L^2 < 0 $, Einstein spaces are solutions
of the equations of motion:
\be
R_{MN} = - 2 L^2 g_{MN}\; .
\label{ein}
\ee
The usual maximally symmetric solution of (\ref{ein})
in $5D$ is $AdS_{4+1}$ spacetime:
\be
ds^2 = -\cosh^2(L r/ \sqrt{2}) dt^2  + dr^2 + (2 / L^2)
\sinh^2 ( Lr/ \sqrt{2})d\Omega_3\; ,
\label{maxads}
\ee
where the metric on $S^3$ can be (up to gauge invariance) taken
as the usual $S^2$ foliation form  $d\Omega_3 =
d\psi^2 + \sin^2 \psi\,\left(d\theta^2 + \sin^2\theta \,d\phi^2\right)$ or
the
Clifford torus ($S^1\times S^1$) foliation form
$d\Omega_3 = d\psi^2 + \cos^2\psi\, d\theta^2 + \sin^2\psi
\,d\phi^2$. Here also notice that, when the cosmological constant is
turned
off, the above metric smoothly becomes the $5 D$ Minkowski metric. But, it
is
clear
that the metric (\ref{maxads}) cannot be deformed to an effective
$3+1 D$ spacetime whose spatial part is $AdS_3$ that supports the
KK monopole with spherically symmetric magnetic field.
In fact, the proper background
should be of the $AdS_2\times AdS_3$ form
with the line element:
\be
ds^2 = - {1\over 2} \exp{(-2\sqrt{2}L x^5)} dt^2 + dx_5^2 +
dr^2 + {1\over L^2} \sinh^2(L r) d\Omega_2\; ,
\label{ads}
\ee
where  $d\Omega_2= d\theta^2 + \sin^2\theta d\phi^2 $.
To make our choice of background
more transparent, let us recall that the background (as well as
the soliton) has to be an Einstein space. The first candidate (
the maximally symmetric one (\ref{maxads}) ), as
argued above, does not work: one cannot compactify
one of the $\psi$, $\theta$ or $\phi$ coordinates on an {\it{arbitrary}}
circle, which is needed for the KK interpretation of the metric.
For the metric (\ref{maxads}) to be Einstein,$\psi, \theta, \phi$
coordinates all take values on {\it{fixed}} radii circles, rendering
the metric too 'rigid' to accommodate a KK monopole.
One, then, has to find another Einstein space. For several other
reasons laid out below, the metric (\ref{ads}) is the suitable one.

Analogous to the flat KK monopole, the
sought-after soliton should yield the above metric in the limit of
zero charge (as
$M\rightarrow 0$). From (\ref{ads}) one can immediately anticipate
the form of the magnetic field
$\vec{B}$. Due to the flux conservation, the magnetic field seen by an
observer who
lives in the effective $3+1 D$ spacetime, should be
\be
\vec{B} = {4 M L^2 \hat{r}\over
\sinh^2{(L r)}}.\;
\label{magnetic}
\ee
This is the field of a $3D$ Abelian `hyperbolic monopole'
\cite{atiyah,chakrabarti} (instead of an $R^3$ monopole of the flat KK
monopole metric). Hence the total magnetic flux $\Phi =: \int \vec{B}\cdot
d\vec{s}$ of the
hyperbolic monopole
(\ref{magnetic})
on a large sphere (for fixed $t, x^5$) is the same as the flat space
case: $\Phi=16\pi M$. With the definition, $B :=  F = dA$,
the gauge potential 1-form retains its flat space form:
$A = 4M\left(1-\cos{ \theta }\right) d\phi$. We
also require that as $\Lambda \rightarrow 0$, the AdS monopole reduces to
the flat KK monopole. This is because, the solution we are
looking for is the `KK-monopole' in a background with negative
cosmological constant, as we turn $\Lambda$ off, we should obtain flat
KK monopole metric (\ref{themetric}).

Bearing these constraints in mind, we
write down the generic, static
ansatz which might solve the equations of motion:
\be
ds^2 = -{1\over 2}a^2(r, x^5)\exp{(-2\sqrt{2}Lx^5)} dt^2 +
b^2(r,x^5)\left[dx^5 +4M\left(1-\cos{\theta}\right) d\phi
\right]^2\nonumber\\
+v^2(r,x^5)\left[dr^2 + (1/ L^2) \sinh^2(L r)d\Omega_2\right]\; .
\label{metric2}
\ee
For the zero magnetic charge limit $(M=0)$, we require $a(r,x^5) =
b(r,x^5)
= v(r,x^5) = 1 $. For $L =0$, the solution should yield:
\be
a^2(r,x^5) = 2  \hskip 1 cm {\rm and} \hskip 1 cm   b^{-1}(r,x^5) =
v(r,x^5) = 1 + {4M\over r} \; .
\ee
We claim that there are no non-trivial monopole-like solutions
of the form (\ref{metric2}) with the desired properties laid out above.
The easiest way to show this, is
to work directly with the equations of motion. We only need to
look at the simplest components of the Ricci tensor, the ones that are
supposed to vanish.
Let us start with $R_{\theta x^5}$:
\be
R_{\theta x^5}=
\frac{
8M^2L^2 b^2(r,x^5)}{v^2(r,x^5)\sinh^2{(Lr)}}
\,\frac{\left(\cos{\theta}
-1\right)}{\sin{\theta}}
\left[\sqrt{2}L-3
\frac{b'(r, x^5)}{b(r, x^5)}-\frac{a'(r, x^5)}{a(r,
x^5)}+\frac{v'(r, x^5)}{v(r, x^5)} \right] \; ,
\ee
where $'$ denotes $\partial_{x^5}$. For non-vanishing $M$ and $L$,
setting $R_{\theta x^5} =0$, we find:
\be
b(r,x^5)= \left[{{v(r,x^5)}\over
{a(r,x^5)}}\right]^{1/3}f(r)\exp{(\sqrt{2}L x^5/3)}\; ,
\label{cond1}
\ee
where $f(r)$ is an arbitrary function. Plugging (\ref{cond1})
into $R_{\theta\phi}$ yields:
\be
R_{\theta \phi} = \frac{4M}{3} {{\left(\cos{\theta}-1\right)^2} \over
{\sin \theta}}
\left[\frac{a'(r,x^5)}{a(r,x^5)} +
2\frac{v'(r,x^5)}{v(r,x^5)}-\sqrt{2}L \right]\; .
\label{cond2}
\ee
Then, the equation $R_{\theta \phi} = 0$ is solved by:
\be
v^2(r,x^5)={g(r)\over {a(r,x^5)}}{\exp{(\sqrt{2}Lx^5)}}\; ,
\label{cond3}
\ee
where again $g(r)$ is arbitrary. Using  (\ref{cond1}, \ref{cond3}),
$R_{rx^5}$ is simplified as:
\be
R_{r x^5}=\frac{3}{2}\frac{\partial_r a(r,x^5)}{a^2(r,x^5)}\left[\sqrt{2}L
a(r,x^5)
-a'(r,x^5)\right]\; .
\ee
For the metric to be Einstein we set $R_{r x^5}=0$,
leading to:
\be
a(r,x^5)&=&h(r)\exp{(\sqrt{2}Lx^5)}\hskip 1cm   {\rm or}\; ,\label{1}\\
a(r,x^5)&=&a(x^5)\; \label{2}.
\ee
Hence, collecting everything, one can write the final metric for the case
(\ref{1}) as:
\be
ds^2=-\frac{1}{2}h^2(r)\,dt^2+\frac{g(r)}{h(r)}\left[ dr^2 + (1/L^2)
\sinh^2(L r)d\Omega_2\right ]\nonumber\\
+\frac{f^2(r)g^{1/3}(r)}{h(r)}\left[ dx^5
+4M\left(1-\cos{ \theta}\right) d\phi\right]^2\; .
\label{metric3}
\ee
Observe that the $x^5$ dependence is completely dropped out. Therefore,
it is clear that $M=0$ does not yield the background metric
(\ref{ads}). The remaining
case (\ref{2}) also fails in recovering the background metric in the
certain limit. Thus, there is no $5 D$ KK monopole in AdS
space that limits to the flat space KK monopole as $\Lambda \rightarrow
0$. Here, although we have shown the non-existence
of KK monopoles in AdS spacetime under some physical requirements in
the corresponding limits, we have
not ruled out
the existence of
any other type of solitons that might live in
cosmological backgrounds.

\section{ 6D AdS monopoles}

In the previous section, we have seen that we cannot embed the
$3D$ hyperbolic monopole in a $5D$ cosmological space. The main problem
was that the time direction could not simply be added to the spatial
part. On the other hand, as we show below, in $D \ge 5+1$
cosmological spacetimes, one can construct solitons which
resemble the flat space KK monopoles.

For monopole solutions, the need for at least two extra
dimensions in the cosmological case should not
be surprising. In fact, the only other non-trivial
soliton we know for $\Lambda<0$ -the AdS soliton,
has to be at least 6-dimensional if 3-dimensional
spherical symmetry is imposed \cite{horowitz}.
Indeed, the KK monopole we are searching for,
is also spherically symmetric. [ Note that in the metric,
one breaks the spherical symmetry by choosing a gauge
for the gauge potential $\vec{A}$. But, physically, the spacetime
is spherically symmetric. ]

To construct a $D= 6$ soliton with a $3D$ magnetic field
of the form (\ref{magnetic}), for the reasons discussed in the previous
section, we take the following
$AdS_2 \times AdS_4$ background:
\be
ds^2 = -\exp{(- 2 L x^6)} dt^2 + dx_6^2
+\frac{3}{2}\left[ dr^2 + (1/L^2) \sinh^2(r L ) d\Omega_2
+ \cosh^2(r L) dx_5^2\right]\; ,
\ee
and deform it to:
\be
ds^2 = -\exp{(- 2 L x^6)} dt^2 + dx_6^2 +\frac{3}{2}H^2(r)
\Bigg\{V(r) \left( dr^2 + (1/L^2) \sinh^2(r L) d\Omega_2\right)
\nonumber\\
+{\cosh^2(r L)\over V(r)}\left[ dx^5 +
4M\left( 1- \cos \theta \right)d\phi\right]^2 \Bigg\}\; .
\ee
This metric is the most general one that meets
our requirements for various limits. The non-trivial
$4D$ part of the metric, with $S^3$ foliations instead of $AdS_3$,
was studied before by Pedersen \cite{pedersen}.
Staticity, `spherical symmetry' and diffeomorphism invariance
leave two independent functions in the metric, best parametrized
as above. Spherical symmetry is obvious in the curvature invariants of the
metric (the only physical quantities), since the non vanishing ones depend
only on $r$. For various
limits, we should get:
\be
M=0  \rightarrow  H(r)= V(r) = 1\; , \hskip 1 cm  \Lambda =0 \rightarrow
H(r)=1\; , \hskip 0.5 cm  V(r) = 1 +{4 M\over r} \; .
\ee
Equations of motion can be simplified by observing that for non-zero
$\Lambda$, $r \rightarrow  \sinh(r L)/L$ in the magnetic field.
Therefore, one can take the ansatz:
\be
V(r) = 1 +{4 M L\over \sinh(r L)}\; .
\ee
Then the remaining function $H(r)$
can be determined from the equations of motion:
\be
H(r) =  {1\over {1 - 4 M L \sinh(r L)}}\; .
\ee
Thus, the $D=6$ AdS KK monopole metric is:
\be
ds^2&=&-\exp{(- 2 L x^6)} dt^2 + dx_6^2\nonumber \\
&&+{{3/2}\over ( 1- 4M L \sinh(r L) )^2}
\Bigg\{\left[ 1+ {{4M L\over \sinh(r L)} }\right]
\left( dr^2 + (1/L^2)\sinh^2(r L)d\Omega_2^2\right) \nonumber \\
&&+\cosh^2(r L)
{\left[ 1+ {4M L \over \sinh(r L)}\right]}^{-1} \left[ dx^5 +
4M\left(1- \cos \theta \right)d\phi\right]^2 \Bigg\}
\label{fullmetric}\; .
\ee
The monopole is located at  $r= 0$. Note that as $r\rightarrow 0$, the
$4D$ part of the metric
reduces to the singular flat TN space given in
(\ref{metric1}). Again, to get rid of the
singularity at the origin, $x^5$ should be compactified on a
circle with radius $8M$.

Because of its direct product structure,
there are two disconnected boundaries of the above metric: One at
$x^6 = - \infty$ and the other at  $L r = \sinh^{-1}( 1/4ML)$. [
$x^6 = + \infty$ is a Killing horizon.] The first boundary
is the usual-trivial boundary of $AdS_2$.
In the $r$-coordinate, the latter boundary is at a finite distance but
can be moved to `infinity' by
transforming the metric into new coordinates $\tilde{r} := \int dr
H(r)\sqrt{V(r)}$. For  $r \in [0, (1/L)\sinh^{-1}( 1/4ML)]$,
$\tilde{r} \in [0, \infty]$.
Only `light' can travel to the boundary at a finite time.
Strictly speaking, for $4 D$ part alone,
there is of course no light-like geodesic. But to understand
the boundary let us set $\phi = \theta = \mbox{const.}$
and set $ds^2_{4D} =0$ to find the Euclidean time:
$ ix^5 = \int_\epsilon^{(1/L)\sinh^{-1}( 1/4ML)}
dr V(r)/\cosh( rL )  = \mbox{finite}$,
where $\epsilon > 0$. [ Note that if the light starts its journey
at $r=0$, it will take an infinite time to reach the boundary, since
$r= 0$ is a Killing horizon.] Thus, the metric (\ref{fullmetric})
describes a non-singular
geodesically complete space. Another way to see the smoothness of the
metric
is to check whether curvature invariants are regular or not.
It suffices to look at the $4 D$ part of the metric.
Among the Riemann invariants \cite{Carminati, Zakhary}, apart
from the obvious ones like the Ricci scalar, $R_{\mu \nu}R^{\mu \nu}$ etc.,
the only non-vanishing invariants are the Weyl scalars:
\be
W_1:\!\!&\!\!\!=\!\!\!&\frac{1}{8}C_{\mu \nu \sigma \rho}
C^{\mu \nu \sigma \rho}=\frac{1}{8}
C^{*}_{\mu \nu \sigma \rho }C^{\mu \nu \sigma \rho}\nonumber\\
&\!\!=\!\!\!&\frac{64M^2\, L^6}{3}\left(\frac{1-4ML\sinh{(Lr)}}{4ML
+\sinh{(Lr)}}\right)^6\; ,
\label{inv1}
\ee
and
\be
W_2:\!\!&\!\!\!=\!\!\!&-\frac{1}{16}C_{\mu \nu }\,^{\sigma \rho
}C_{\sigma
\rho }\,
^{\lambda \eta }C_{\lambda \eta }\,^{\mu \nu}=
-\frac{1}{16}C^{*}_{\mu \nu}\,^{\sigma \rho }C_{\sigma \rho }\,
^{\tau \eta }C_{\tau \eta }\,^{\mu \nu}\nonumber\\
&\!\!=\!\!\!&\frac{512\,
M^3\,L^9}{9}\left(\frac{-1+4ML\sinh{(Lr)}}{4ML+\sinh{(Lr)}}\right)^9\;
, \label{inv2} \ee where $C_{\mu\nu\sigma\rho}$ is the Weyl tensor
and $C^*_{\mu\nu\sigma\rho} :=
\frac{1}{2}\epsilon_{\mu\nu\lambda\eta}
C^{\lambda\eta}\,_{\sigma\rho}$ is its dual. None of the
invariants are singular. Weyl scalars vanish at the infinity where
$L r = \sinh^{-1}( 1/4ML)$ in our coordinates. As a check of the
calculation of the above invariants, let us note that
$\Lambda\rightarrow 0$ limit of the $4 D$ part of
(\ref{fullmetric}) is the flat TN metric multiplied by $3/2$ and
has the following Weyl scalars: \be
W_1&=&\frac{64}{3}\frac{M^2}{(4M+r)^6}\; ,\\
W_2&=&-\frac{512}{9}\frac{M^3}{(4M+r)^9}\; ,
\ee
which are nothing but the $\Lambda \rightarrow 0$
limits of (\ref{inv1}) and (\ref{inv2}) respectively.

Finally, let us say a few words about
the energy of the metric ({\ref{fullmetric}}).
The gravitational energy of the flat space KK monopole
was studied in \cite{sorkin2, soldate}. As is often the case
for the gravitational systems with compact dimensions,
a useful energy definition is rather tricky because of the
complications in choosing a proper background metric with
respect to which the energy should be defined.
The total gravitational energy of a given metric makes sense,
if it can be considered as a `particle'
or a `perturbation' in a given background or vacuum. The
background, by definition, solves
the equations of motion everywhere and has zero energy.
The topology of the background metric ought to be the same as the
metric whose energy is measured.
The flat space KK monopole is a perfectly smooth
solution, and there is no other one with the same topology
whose energy can be compared with the monopole.
Therefore the KK monopole, considered  as {\it{the}} vacuum,
has trivially zero energy.
We should note that in \cite{sorkin2,soldate}, the
chosen background metrics are {\it{not}} solutions everywhere.
They are {\it{only}} asymptotic solutions.
Our metric ({\ref{fullmetric}}), being a smooth solution everywhere,
has also zero energy. This being a rather
boring result, one might, like
\cite{sorkin2,soldate}, relax the condition on the background and
take an asymptotic solution as a background. Here, we do not go into
the details but heuristically argue that the AdS KK monopole's
energy
expression is similar to the flat space one.
Following Abbott and Deser (AD) \cite{abbott}~\footnote{See also {\cite{dt}}},
let us choose $\bar{g}_{\mu \nu}$ as the background for which $V(r) =1$
in (\ref{fullmetric}),  and  $h_{\mu \nu}$
as the perturbation part, defined as  $h_{\mu \nu} :=  g_{\mu \nu}
- \bar{g}_{\mu \nu}$. We do not drop the $g_{x^5\phi}$ term: namely, the
vector potential is taken to be {\it{non}}-zero, to keep the topology
of the background the same as the KK monopole.
Then, $O(h)$ terms in the equations of motion
constitute the background-conserved energy momentum tensor
\be
R_{\mu \nu}^L - {1\over 2} \bar{g}_{\mu \nu} R^L - 2\Lambda h_{\mu \nu}
:= T_{\mu \nu } (h) \; ,
\ee
where $L$ denotes linearization. Given ${\bar{\xi}^{\mu}}$ as a
background Killing vector, an ordinarily conserved current
can be constructed: $\bar{\nabla}_{\mu}( \sqrt{-\bar{g}}\, T^{\mu \nu }
\bar{\xi}_\nu ) =
\partial_{\mu}( \sqrt{-\bar{g}}\, T^{\mu \nu }\bar{\xi}_\nu )$.
Using this, we write the AD Killing energy, up to trivial
numerical factors, as
$E= \int d^5 x \sqrt{-\bar{g}}\,T^{0 \nu}\bar{\xi}_\nu $. For
the metric (\ref{fullmetric}), $\bar{\xi}^\nu = (-1, {\vec{\bf{0}}})$,
thus $E =  \int d^5 x \sqrt{-\bar{g}}\, T^{0 0}\bar{\xi}_0 $. Since
the AdS KK monopole is an Einstein space,
$R^L = 0 \rightarrow R_{\mu \nu}^L = 2 \Lambda h_{\mu \nu}$ and hence
$T^{00} = 0$ everywhere, except at $r=0$ where the derivatives give
$\delta$-function singularities. Therefore, all the contributions to
the energy volume integral come from the origin. We already
know that the non-trivial part of the metric $(\ref{fullmetric})$
reduces to the flat-space KK monopole metric around the origin. Thus,
the AdS KK monopole
has the same energy expression as the flat one, which was
obtained to be $M$ in \cite{sorkin2, soldate}. The above argument
is heuristic and depended on the `volume' integral of the
energy expression. One could, in principle, turn the volume integrals
for conserved charges (like energy) into surface integrals at infinity.
This method should yield, albeit after a lengthy calculation,
the same result as above.

{\bf{ A 6D KK Monopole Brane-World}}

Having six dimensions at our disposal, we can construct more
than one kind of solitons. Here, we relax our earlier condition that
the low energy spacetime has a negative cosmological constant.
Using the five dimensional flat Kaluza-Klein monopole, a six dimensional
soliton can be given by the following line element:
\be
ds^2= \exp{( 2 k L x^6)}\left \{-dt^2+ ds^2_{\mbox{{\scriptsize TN}}}
\right \} +dx_6^2\; ,
\label{example1}
\ee
which satisfies $R_{MN}=-5 k^2L^2g_{MN}$.
Multi-monopoles with equal charges, located at
different spatial points can be obtained as in the flat space case.
$x^5$ direction  has to be compact for the solution to be smooth but
the $x^6$ direction has an infinite extent. Generalization to  $D$
dimensions is straightforward: one can add more flat directions
as long as they are multiplied by $\exp{(2k Lx^6)}$.

Let us choose $k=-1$ for definiteness. Then, $x^6 = -\infty$ is
the boundary of the space and $x^6 = + \infty $ is the Killing
horizon. In this form, all dimensions (except the compact $x^5$
directions) are accessible to the low energy observers.  To
localize the gravity, and get an effective $3+1 D$ brane-world, we
can cut and paste (\ref{example1}) into the Randall-Sundrum $Z_2$
invariant form \cite{randall}: \be ds^2=\exp{(- 2 L|x^6|)}\left
\{-dt^2+ ds^2_{\mbox{{\scriptsize TN}}} \right \} +dx_6^2\; .
\label{example2} \ee This metric is not differentiable at the
`origin', where the brane ($D4$ Taub-NUT brane ) is located.
Physically, $3+1 D$ observers see a flat-space KK monopole. The
metric ({\ref{example2}) is a six dimensional brane world with
gravity localized on a Ricci-flat brane.

\section{Conclusions}

We have shown that the five dimensional cosmological
Einstein gravity (with a Lorentzian
signature and a negative cosmological constant ) does not have Kaluza-Klein
monopole type static soliton solutions. On the other hand, in $D \ge 6$,
we have constructed analogs of the flat space KK monopoles which are
asymptotically locally AdS space-times. It would be interesting
to find out if these solutions preserve some supersymmetry and if
they can be embedded in string or M-theory. In the presence of
anti-symmetric tensor fields, in addition to the usual translational
collective degrees of freedom, the flat space KK monopole
($D5$ or $D6$ brane ) can
have dyonic deformations \cite{sen}.
An extension of these ideas to the AdS KK monopole would also be worth
studying.

In this paper, we have insisted the solitons to
be static. If we relax this condition,
we can find  time-dependent $4+1 D$ or Euclidean $5D$ solutions.
One such (Euclidean) example with the Ricci-flat 4D slices would be:
\be
ds^2 =  dt^2 + \exp{(- 2tL)}\, ds^2_{\mbox{{\scriptsize TN}}}\; .
\ee
Another example, with negatively curved $4D$ slices is:
\be
ds^2 = - dt^2+{\sin^2(t L)\over ( 1- 4M L \sinh(r L) )^2}
\Bigg\{\left[ 1+ {{4M L\over \sinh(r L)} }\right]
\left( dr^2 + (1/L^2)\sinh^2(r L)d\Omega_2^2\right) \nonumber \\
+\cosh^2(r L)
{\left[ 1+ {4M L \over \sinh(r L)}\right]}^{-1} \left[ dx^5 +
4M\left(1- \cos \theta \right)d\phi\right]^2 \Bigg\}\; .
\ee

Construction of multi-monopoles in AdS, which does not seem to
be a straightforward task, is under current investigation.

{\bf{Acknowledgments}}

Authors would like to thank S. Deser, A. Lawrence, P. Ramond,
P. Sikivie, R. P. Woodard and N. Wyllard for useful discussions,
encouragement and support. The work of B.T.
is supported by NSF grant PHY99-73935. V. K. O. is supported by
DOE grant DE-FG02-97ER41029.

\vskip 1cm

%\leftline{\bf References}

\myend